\documentclass[a4paper,UKenglish]{lipics}
 
\usepackage{microtype}

\title{Logic Modelling\footnote{The author gratefully acknowledge the financial support of the Natural Sciences and Engineering Research Council of Canada (NSERC).}}

\author[1]{Roger Villemaire}
\affil[1]{Department of Computer Science\\
  Universit\'e du Qu\'ebec \`a Montr\'eal\\
  C.P.\ 8888, Succ.\ Centre-ville,\\
  Montr\'eal, Qu\'ebec, H3C~3P8, Canada\\
  \nolinkurl{villemaire.roger@uqam.ca}}
\authorrunning{R. Villemaire} 

\Copyright{Roger Villemaire}


\serieslogo{logo_ttl}
\volumeinfo
  {M. Antonia {Huertas}, Jo\~ao {Marcos}, Mar\'ia {Manzano}, Sophie {Pinchinat}, \\
  Fran\c{c}ois {Schwarzentruber}}
  {5}
  {4th International Conference on Tools for Teaching Logic}
  {1}
  {1}
  {263}
\EventShortName{TTL2015}

\begin{document}

\maketitle

\begin{abstract}
  This is a reflection on the author's experience in teaching logic at the graduate level in a computer science department. The main lesson is that model building and the process of modelling must be placed at the centre stage of logic teaching. Furthermore, effective use must be supported with adequate tools. Finally, logic is the methodology underlying many applications, it is hence paramount to pass on its principles, methods and concepts to computer science audiences.
 \end{abstract}

\section{Introduction}
Academically, computer science emerged from engineering and mathematics propelled by the development of computing hardware. While logic is clearly identified with the theoretical foundation of computing, it also plays an important role in many applications. Unfortunately while its method are central to many technologies, logic goes almost unnoticed for most computer scientists.

Still, there are logic courses in computer science departments, at least at my university. But the question emerges as to what to teach and how. I present in this paper three graduates logic courses that I developed over the years. Each has different objectives and aim at a different audience. Nevertheless, all three have in common to present applications and to develop the students' abilities to effectively use the material. They also illustrate how modelling is at the heart of logic, as it is used in computing.

In order to present my approach, I will briefly describe the courses' contents in Section \ref{SECCP} before presenting in Section \ref{SECLOG} a unifying view for applications of logic to computer science and showing that my courses are indeed structured that way. I will also describe surprises and difficulties that the students usually encounter. Finally, in Section \ref{SECCON}, I will conclude on my teaching experience.

Logic offers a quite general methodology and its ideas, methods and algorithms underlie many important applications. But this goes almost unnoticed, holding up the development of new applications. This paper presents my approach to reaching out to graduate computer science students.
\section{Courses presentation}\label{SECCP}
This section briefly presents the three courses considered in this paper. All are graduate courses, but with different contents and quite different audiences. All courses are 15 weeks long, with the class meeting once a week for a 3 hours lecture. Between a third and the half of these 45 contact hours are spent on exercises and effective use of the tools. While each student is encouraged to come up with his own solutions, groups are usually small (between 4 and 12 students) and there is therefore a lot of interaction between students and with the lecturer.

A word on the students' background and their previous exposure to logic and the fundamentals of computer science is in order. For the computer science and software engineering classes presented in subsections \ref{SUBSECMV} and \ref{SUBSECFM}, while students previously attended usually many different undergraduate institutions, they all have a basic training in discrete mathematics and algorithmic. As for logic itself, they have usually learned some propositional and first-order logic, but mainly as a formalism to express properties with no reference to inference and deduction systems. As for the course presented in subsection \ref{SUBSECLICS}, the audience's background is broad, ranging from computer science to the humanities. While the students with a computer science background have a training similar to the students of the two previous courses, this is not the case for students from the humanities. Nevertheless, this last group of students have usually been exposed to some form of formal systems. This is particularly true for students with a background in linguistic, psychology and philosophy, which form the major part of the humanities students in this class.

\subsection{Modelling and Verification}\label{SUBSECMV}
This course covers symbolic model-checking \cite{cemgopdaMIT00} and is an optional course in the master's and PhD in computer science, both being traditional research oriented degrees.

The first half of the course is devoted to modelling and the second half to covering the algorithms and data structures allowing efficient verification. 

Modelling was done up to the last term with the NuSMV tool \cite{acemcegfgmpmrrsat02,NuSMVman24} and now with its follow-up NuXMV \cite{NuXMVman10}. Without going into the details of the tool's language, a NuSMV/NuXMV model defines variables that range over finite domains and describes variables' evolution. Initial values are assigned to variables and the (discrete) evolution of the system is given by specifying \emph{next} values of variables. The underlying model is hence a finite state machine, where a \emph{state} is determined by the value of variables at some moment and the \emph{next state} is determined by the next values of the variables. The tool supports non-deterministic behaviour by allowing more than one initial and next values for variables. The tool determines the validity of \emph{Linear Temporal Logic} (LTL) and \emph{Computation Tree Logic} (CTL) formulas on a given model. The reader is referred to \cite{eaeHTCSB_1990} for an introduction to temporal logic.

The algorithmic methods presented in the course (and supported by the tool) are \emph{Bounded Model-Checking} (BMC) \cite{abacemcyzTACAS99}, which is based on efficient SAT-solvers, and BDD-based model-checking \cite{jrbemcklmdldljhIC_98_2_1992}. Detailed lecture notes (in French) have been produced \cite{rvMV2013} and are freely available under a Creative Common license.


\subsection{Formal Methods}\label{SUBSECFM}
This course covers formal methods in software engineering and is an optional course in the master's degree in software engineering, a professional degree with mostly part-time students.

Software modelling is done in the \emph{Unified Modeling Language} (UML) a standard of the \emph{Object Management Group} (OMG)\footnote{\href{http://www.omg.org}{\nolinkurl{www.omg.org}}}. Since this is the standard in the profession, the students usually already master this language and have at least basic modelling skills. The course introduces logic as a tool to refine and make models more detailed and precise. Two applications are presented. First test generation and then software verification. Finally, sequent calculus is introduced as this is the method used by the software verification tool (see Section \ref{SECFM}). This course follows a ``hands-on'' approach, where each class combines concepts and notions that are immediately put in application.

The logics presented are the \emph{Object Constraint Language} (OCL), a part of the UML standard, and \emph{Java Modeling Language} (JML), that is specifically designed for Java programs. Modelling in done with the \emph{Eclipse Modeling Tools}\footnote{\href{http://www.eclipse.org/modeling/}{\nolinkurl{www.eclipse.org/modeling/}}}. An Eclipse plug-in, \emph{Dresden OCL}\footnote{\href{http://www.dresden-ocl.org/}{\nolinkurl{www.dresden-ocl.org/}}}, is used to help writing OCL expressions and to generate tests. Finally, the \emph{KeY}\footnote{\href{http://www.key-project.org}{\nolinkurl{www.key-project.org}}} \emph{Automatic Theorem Prover} (ATP) is used to verify JML properties on Java code. Slides, examples, and course material are available on the course's Web site (in French)\footnote{\href{http://moka.labunix.uqam.ca/~villemaire_r/7160.html}{\nolinkurl{moka.labunix.uqam.ca/~villemaire_r/7160.html}}}.

\subsection{Logic, Informatics and Cognitive Sciences}\label{SUBSECLICS}
This course covers logic as a tool for human cognition and knowledge representation. It is an optional course for students in the \emph{Cognitive Informatics Doctorate}, a joint program between the computer science department and the faculty of humanities. Following the tradition in this program, the course is given by two lecturers, namely, my colleague Serge Robert, a logician from the philosophy department, and myself.

After the first lecture where we present the course, its objectives, and our approach, and we ask the students about their expectations, we each take care of half of the lectures. Serge covers topics such as classical, non-monotonic, many-valued, and probabilistic logics in relation to human cognition. For my part, I concentrate on knowledge representation with modal and description logics. I introduce tableaux as an inference method and allow time in each class to let the students represent some knowledge and use a tool to experiment with inference.

For modal logic, I use LoTREC \cite{lfdcogahmsICTTL06}, in fact a slight modified version\footnote{LoTREC 2.1 \href{http://moka.labunix.uqam.ca/~villemaire_r/LoTREC/}{\nolinkurl{moka.labunix.uqam.ca/~villemaire_r/LoTREC/}}}. This tool allows me to attain two different, but fundamental, objectives. First, the student can graphically visualise tableau proofs and hence verify that she/he correctly masters the algorithm. This is particularly useful for students of the humanities that can be quite intimidated by algorithmic contents. Secondly, the tool allows me to put tableaux at work to infer new knowledge. Students can hence experiment and determine shortcomings of their tentative representations. This is particularly important when faced with the choice of the appropriate modal logic, such as K4, S4, or S5.

For description logic, students are introduced to devising ontologies in Prot\'eg\'e\footnote{\href{http://protege.stanford.edu}{\nolinkurl{protege.stanford.edu}}}. Within this tool, Hermit\footnote{\href{http://hermit-reasoner.com}{\nolinkurl{hermit-reasoner.com}}} is used to derive the inferred hierarchy. This allows the student to test whether her/his ontology is free of contradictions and, more importantly, to see in action how tableaux can infer derived knowledge.

Slides, examples, and course material are available on the course's Web site (in French)\footnote{\href{http://moka.labunix.uqam.ca/~villemaire_r/9305.html}{\nolinkurl{moka.labunix.uqam.ca/~villemaire_r/9305.html}}}.
\section{Logic: Information, Representation and Processing}\label{SECLOG} What is logic and how can it be presented to a computer science audience? I present in this section a view of logic around information processing, describe the material from my courses within this unifying perspective, and, finally, discuss the typical difficulties and surprises students encounter during my lectures.
\subsection{Information} Logic is a tool to represent and process information. While computing, a.k.a. informatics, is the science of information, it should be stressed that information goes much further than what is usually considered as data. Indeed, logic offers quite a large range of abstraction levels. One can, as in databases, speak of a specific \emph{person $P$ having the person $M$ as mother}, but one can also consider the \emph{motherhood} relation and state that \emph{every person has a mother}. Information has content and meaning, and it is used for some purpose. Obviously, quantity and relevance also matters: one cannot conclude if there is not enough information or if it is not relevant. Information must be stated in some way, this is what logic is about.
\subsection{Models and Properties}
Logic offers formalisms to represent information\footnote{We limit ourselves to \emph{formal} logic in this paper.}. Being formal means that there is a precise syntax and semantics, much less ambiguous than for natural languages and more easily machine processed. Logic is hence a general approach to knowledge representation, allowing many levels of abstraction. Essentially, any knowledge can be represented, but the exact formalism and formalisation must be chosen with care, keeping in mind the intended goal. Logical formalisms hence make precise, concepts, ideas, relations, and data.

Aiming at applications to computing, a logic description can be presented as a pair formed of a \emph{model} and a \emph{property}. Technically, this distinction is artificial, since all pieces of information could be defined either all within the model or all within the property. But this distinction is convenient and natural.

The distinction between model and property lies in fact in the intention. The model is intended as a (partial) description of the system. We borrow the notion of \emph{system} directly from the practice of computer science. It is (the formalisation of) the entity one wants to interact with. One can think of a database, a software module, a communication protocol, the description of some process, a knowledge base, etc. A model is some, possibly partial, view of some system of interest. By contrast, a property is a constraint (on the model) that one wants to check or verify. For their part, properties are naturally partial; we are not describing everything that should hold, but something that should hold.

Model and property are closely bound to one another. The model is meaningful for a specific goal, (essentially) formalised by the property. Conversely, the property is checked on the model. It is hence a constraint on the model's elements or \emph{attributes} as they are called in software engineering. As modelling is a creative endeavour, there is no single right model or property; model and property must be appropriate for the intended goal.

To sum up, from the perspective of a computer scientist making effective use of logic, the model describes a system whose significance transcends the verification task, while the property is confined at this task's heart, as it expresses what will be verified.

This distinction, natural from a computing point-of-view, applies nicely to the material presented in the three courses.

\subsubsection{Modelling and Verification}
In this class, the model is represented by a \emph{finite state machine}. One defines (finitely many) variables ranging over finite data types and assigns initial and next values to variables. As the model's evolution is described in discrete time, the property is stated in LTL or CTL.

This modelling process can be naturally introduced to computer scientists, since at the graduate level they are already quite experienced in software development. A piece a code is indeed a formal model, but of an extremely detailed form. As for the property, which seems at first less natural from a computing point-of-view, students have encountered assertions and testing in software development and checking a property is hardly something new.

While this analogy helps introducing the concepts and get the student engaged with the material, it only goes so far. It is important to emphasise that the objective in formal verification is to check that the model fulfil the property and not to implement and debug a functional system. This comes at some surprise to mainly development-oriented students and necessitates a throughout discussion supported by hands-on modelling and verification experience during classes. Furthermore, contrary to testing, the objective it to verify the property for \emph{all behaviours} of the model. This will offer some difficulties.

The students' natural tendency is to develop overly detailed models. A main objective of the course is to make the students aware that this is neither needed nor appropriate. Indeed, the goal is to verify a property which usually depends only on a limited number of details. Furthermore, a simpler model will usually mean a more efficient verification. Finally, a simpler model will be easier to develop (or generate) and maintain, a consideration natural for computer science students.

A nice simplification example is offered by a scheduler that guarantees, at each moment, the execution of a unique component. There may be some ``real'' deterministic scheduler implementing some complex algorithm. But if the objective is not to show the scheduler's correctness, but rather that of illustrating a more complex system using a scheduler, non-determinism can largely simplify the model. Indeed, in this case the model can simply specify that the scheduler always chooses a unique component without having to detail how this choice is done.

In developing models during the class, students are also led to realise that models can be of quite different nature. For instance, in modelling a network one can either models packet flow or else abstract away and concentrates on the dependencies between the network equipments' configurations. The real question being always that the model should be adequate to the verification task.

As surprising as it may seem, students have not major difficulties writing LTL or CTL formulas. This may come from the fact that in model checking the model is usually complex and the formula simple. The difficulties usually come from the verification process. Indeed, a slight mismatch between the model and the formula will often lead the tool to return an unexpected counterexample. Often this counter-example hints in fact to the subtle difference between the intended and the actual formalisation. As the students can be rapidly overwhelmed by the subtlety uncovered by the counter-example, it is paramount to teach a progressive and systematic way to analyse counter-examples. 
\subsubsection{Formal Methods}\label{SECFM}
This is a course of a software engineering program. Accordingly, the models considered are cornerstones of the profession, namely, UML class diagrams and Java code. Simply speaking, a UML class diagram describes classes (set of objects), (binary) relations between classes, and, finally, operations, which are function names with arguments, return value, and their types (classes). A UML class diagram does not describe the system's behaviour as there is no provision to describe how operations should behave. In order to consider software behaviour, this course relies on the programming language Java.

For properties, OCL, a not so well-known part of the UML standard, and the dynamic logic of the KeY tool are used. Some explanations about these formalisms are in order.

In a class diagram, relations are binary, but can be of any kind: one-to-one, one-to-many, many-to-one, or many-to-many. Essentially, OCL allows one to constrain relation extremities and in the case of a ``many'' extremity, its size. For example, a \texttt{hasPassed} relation between a student class and a course class would be many-to-many, as each student can have passed many courses, and each course can have been passed by many students. A possible OCL constraint would be to state that every (graduate) student must have passed $30$ (undergraduate) courses. Is some way, OCL allows one to write constraints reminiscent of (multi-)modal logic or description logic.

This course also uses the KeY tool's logic to verify Java code. This is a dynamic logic built from atomic formulas (comparisons between variables and values), propositional connectives and modal connectives $\langle a\rangle$, $[a]$, where $a$ is a Java program. Here $\langle a\rangle\varphi$ means that $a$ will terminate normally (no exception will be raised) and then $\varphi$ will hold, while $[a]\varphi$ means that if $a$ terminates normally, then $\varphi$ holds.

In this course, students have little difficulties with models, since they are already trained in devising UML class diagrams and developing Java code. On the opposite, writing OCL constraints is quite challenging. But the benefit is that it coerces them to take UML modelling seriously. Indeed, in typical software engineering courses, software design is done by adding constraints as comments to UML diagrams. Documentation in natural languages can be pretty imprecise and its correctness tedious to assure. Students are hence not incited toward rigour in UML modelling. On the opposite, writing up OCL constraints forces students to get committed and make their ideas more precise. It also offers, as students appreciate, machine support.

Indeed, OCL being formal, the tool, in our case Dresden OCL, detects such errors as type mismatch. As simple as this may seem, it has a considerable impact. Devising an OCL constraint imposes to navigate the class diagram to access the data. Something as simple as incorrect type detection can already come a long way to notice incorrect navigation. Finally, Dresden OCL allows one to generate Java tests from OCL constraints, showing again the benefit of machine processing of formal languages.

For code verification, the course introduces JML, which allows one to state class invariants and contracts in the form of pre/post conditions on operations. Pre and post conditions are quite natural for students, since they are already largely used to document code. Invariants are more challenging, since they define what a correct instance of a class must be, a modelling task nevertheless completely in line with the course.

The KeY tool is an automatic theorem prover for JML contracts (under the fact that invariants hold). The tool implements a sequent deduction system for its dynamic logic. Students learn sequent proofs quite rapidly, since it is a rule based system. The greatest difficulty is coping with inconclusive proofs. In such a case, I engage students to be attentive to proof branching and unjustified sequents (in particular their meanings), in the search for a missing piece of information. For instance, if an open sequent premise contains $x=null$, one should consider whether $x\not=null$ shouldn't be added to preconditions. Unfortunately, proof analysis is challenging since one must master the formalised Java execution model, in particular the quite complex memory model.

Students find this last part quite challenging and they are right that it considerably increases the burden of code development. But, it must be said that this material is presented as an advanced method appropriate when expectation is high, such as for safety critical code.

\subsubsection{Logic, Informatics and Cognitive Sciences}
In this course both the model and the property are formalised within a single logic. The model is intended as the representation of some knowledge, while the property is a query on this knowledge. One hence wants to determine whether the property is a consequence of the knowledge or not. The logics covered are modal logic and its extension, description logic.

Modal logic extends propositional logic with two unary connectors $\Box$ and $\Diamond$. Under Kripke semantics, the connectors $\Box$ and $\Diamond$ are interpreted via a binary relation with the intuitive meaning ``for all neighbours'' and ``there is a neighbour'', respectively. Modal logic is used to represent facts about necessity/possibility, time, and knowledge. Description logic is an extension of modal logic allowing multiple binary relations and is used in the context of the semantic web to describe and process ontologies, which are, broadly speaking, structured terminologies.

Students in this course come partly from computer science and partly from the humanities. Emphasis is not on theoretical and algorithmic treatments, but rather on the meaning that we want to convey through modelling. Semantics is paramount and the meanings of various typical formulas are analysed, compared, and discussed. Much time is spent on representing knowledge, exploring and comparing alternatives. But such a course could not be complete without explaining how inference can be mechanised. Tableaux are introduced as a common inference framework, first for propositional logic, then for modal logic and finally, in outline, for description logic.

An ``information flow'' approach to modelling is put forward in this course. Reflecting on the model and property, it is emphasised that the model should formalise enough knowledge to be able to draw a conclusion on the property. It is furthermore emphasised that many conclusions can be inferred from the model, but no more than follow from its intuitive meaning. The completeness of tableaux systems is proved, bridging the gap between the intended semantics and algorithmic procedures. It often comes quite as a surprise, for students, that meaning can be processed in such a general way. The significance of completeness is furthermore stressed in concrete situations by modelling significant knowledge. It is worth noting that as tableaux form a deduction system operating on the formulas structures, it falls well into this ``information flow'' approach.

Students are largely stunned by logic. Many students of the humanities are at first disconcerted that a pretty general formalism, such as modal logic, can model possibility/necessity, time, and knowledge. They would rather expect a specific formalism for each situation. It is hence relevant to put logic forward as a universal approach to knowledge representation, similarly to arithmetic that is an adequate formalism for counting things of all nature.

Computer science students have also their own surprises. While they are typically exposed as undergraduates to propositional and first-order logic as representation systems, they have rarely encountered deduction and hence the power of inference in logic. This is particularly significant with description logic. Many have been introduced to the semantic web, but, as usual in that field, mostly as a set of technologies with emphasis on encodings and standards. It is hence a surprise for them that description logic in not only a knowledge representation framework, put also an inference framework allowing to draw up new facts.

\section{Conclusion}\label{SECCON}
When I started to teach logic to computer science audiences, almost twenty-five years ago, my now retired colleague Lorne Bouchard advised me to find the proper tools, if I really wanted to engage the student in the subject. I hence always try to find a tool that is efficient, but also usable in practice to handle significant problems. I some sort of way, I don't look for a tool for teaching logic, but more to the opposite, for a tool that can be used in practice and that uses some form of logic.

This means that I build a course around effective modelling and the use of the tool to process a model toward some useful goal. But, particularly at the graduate level, the students must be exposed to the conceptual machinery that allows such tools in the first place. I hence always introduce the concepts and methods of logic, with a level of justification appropriate for the audience.

There are more and more efficient computing tools powered by ideas, methods, and algorithms from logic. This opens up the way to make computer scientists more aware not only of the power of logic, but also of its methods. There is a logic methodology, based on modelling and powered by efficient engines, that should be much better known in computing. Ironically, there are more and more applications of logic to computer science, but unfortunately still inadequate exposure to our field. I hope to see, in the future, many more applications, but also the return of logic to the central stage of computer science teaching. Logic has a message, something particular to convey, that should properly reach computer science students.

\subparagraph*{Acknowledgments}
I would like to thank the anonymous reviewers for numerous comments and suggestions that have improved the final version of this paper.








\newpage
\thispagestyle{empty}
{\ }

\end{document}